%% file: paper.tex
\documentclass[journal]{IEEEtran}
\usepackage{flushend}
\usepackage{cite}    
\usepackage{comment}    
\usepackage{graphicx}  
\usepackage{caption}
\usepackage{subcaption}
\usepackage{caption}

\usepackage[table,dvipsnames]{xcolor}  
\usepackage{amssymb}
\usepackage{mathrsfs}
\usepackage{mathtools}   

\usepackage[normalem]{ulem}
\usepackage{cancel}

\usepackage{hhline}
\usepackage{tabularx,colortbl}

\usepackage{comment}

\usepackage{multirow}
\usepackage{booktabs}
\usepackage[noend]{algpseudocode}
\usepackage{algorithm}

\newtheorem{assumption}{Assumption}

\def\CC{\mathbb C}
\def\RR{\mathbb R}

\def\cG{\mathcal G}
\def\cI{\mathcal I}

\def\cM{\mathcal M}
\def\cN{\mathcal N}

\def\cT{\mathcal T}

\newcommand{\hide}[1]{}
\newcommand{\raf}[1]{(\ref{#1})}

\newcommand{\re}{\ensuremath{\mathrm{Re}}}
\newcommand{\im}{\ensuremath{\mathrm{Im}}}

\newcommand{\cV}{\ensuremath{\mathcal{V}}}
\newcommand{\cE}{\ensuremath{\mathcal{E}} }

\newcommand{\Opt}{\ensuremath{\textsc{Opt}}}

\DeclareCaptionType{copyrightbox}

\newtheorem{innercustomthm}{\textbf{Theorem}}
\newenvironment{customthm}[1]
  {\renewcommand\theinnercustomthm{#1}\innercustomthm}
  {\endinnercustomthm}

  \newtheorem{innercustomlem}{\textbf{Lemma}}
\newenvironment{customlem}[1]
  {\renewcommand\theinnercustomlem{#1}\innercustomlem}
  {\endinnercustomthm}

\title{Online Algorithm for Demand Response with Inelastic Demands and Apparent Power Constraint}

\author{
Areg Karapetyan, Majid Khonji, Chi-Kin Chau, and Khaled Elbassioni 
\thanks{A. Karapetyan, M. Khonji, C.-K. Chau, and K. Elbassioni are with the Dept. of EECS at Masdar Institute of Science and Technology, UAE (e-mail: \{akarapetyan, mkhonji, ckchau, kelbassioni\}@masdar.ac.ae).}
}

\begin{document}
	\setlength{\abovedisplayskip}{5pt}
	\setlength{\belowdisplayskip}{5pt}
	\setlength{\abovedisplayshortskip}{5pt}
	\setlength{\belowdisplayshortskip}{5pt}

\maketitle

\begin{abstract}
	\input{abstract}
\end{abstract}
{\keywords Demand response management, inelastic demands, apparent power constraint, competitive online algorithm.}

\input{intro}
\input{related}

\input{model}

\input{alg-greedy}
\input{voltage}

\input{sims}
\input{concl}

 \input{append}

\bibliographystyle{ieeetr}
\bibliography{reference}


\end{document}

%% file: abstract.tex
A classical problem in power systems is to allocate in-coming (elastic or inelastic) demands without violating the operating constraints of electric networks in an online fashion. Although online decision problems have been well-studied in the literature, a unique challenge arising in power systems is the presence of non-linear constraints, a departure from the traditional settings. A particular example is the capacity constraint of apparent power, which gives rise to a quadratic constraint, rather than typical linear constraints. In this paper, we present a competitive randomized online algorithm for deciding whether a sequence of inelastic demands can be allocated for the requested intervals, subject to the total satisfiable apparent power within a time-varying capacity constraint. We also consider an alternative setting with nodal voltage constraint, using a variant of the online algorithm. Finally, simulation studies are provided to evaluate the algorithms empirically.

%% file: intro.tex
\section{Introduction}

Today's smart grid requires timely control decision-making in dynamic environments, while ensuring the robustness of electric networks. Recently, there have been numerous studies focusing on viable demand response (DR) management. However, there is limited attention devoted to inelastic demand management and real-time control decisions. The control operations of customers' demands in practice usually consist of binary decisions (e.g., appliances that can be either switched on or off). These demands are often requested in an ad hoc manner, without planning. Hence, control decisions are computed without the knowledge of future information. 

Typically, in a DR management scheme, there is a single load-serving entity (LSE) or an operator of microgrid (MG), who coordinates the scheduling of DR participant customers' demands over a certain time horizon. Various approaches for modeling the energy management in MGs have been proposed in the literature with different extents of consideration of characteristics and operating constraints. The system under study distinguishes between two distinct settings which are driven by the corresponding operating modes of an MG, namely grid-connected and islanded. These settings feature several realistic characteristics of power systems, such as non-linear apparent power constraint and nodal voltage constraint.

There is a high probability that an MG once operating in isolated mode will be short of power. The first setting resembles such a scenario, where customers may suffer from a reduction of generation occasionally due to limited MG capacity. The employed MG model encompasses a hybrid mix of traditional and renewable energy (RE) supplies that could collectively have a variable (depending on the availability of RE and storage available) yet dispatchable capacity. The intermittent RE sources induce time-varying generation capacity. Constrained by the generation fluctuating over time, LSE is required to make binary control decisions in real time as to maximize the total utility of satisfied customers. To solve the resulting optimization problem, a competitive randomized online algorithm is proposed with a definite theoretical guarantee on the ratio over the offline optimal solution. The algorithm relies on the primal-dual schema introduced in \cite{TCS-024}. 

On the other hand, when interconnected to the main grid MG may exchange energy with it, consequently provisioning sufficient generation supply to meet the demand for increasing customer participation. With growing number of customers, however, the load on MG may bring in significant voltage deviations. This is captured by the second setting considered, where LSE makes online binary control decisions to maximize the cumulative utility of satisfied customers subject to the nodal voltages of the distribution network not exceeding the acceptable range. The voltage constrained problem is studied under a relaxed model assuming path topology and small transmission power loss. A variant of the online algorithm is invoked to solve this problem.

%% file: related.tex
\section{Related work}

\subsection{Online Algorithms}

Online algorithms are critical for a wide range of problems involving uncertain input and timely decisions \cite{Borodin:1998, Lian13, CZC16ESP, CKA16DIA}. In an online problem, a sequence of input is revealed gradually over time. The algorithm needs to make certain decisions and generates output instantaneously over time, based on only the part of the input that have seen so far, without knowing the rest of the input in the future.  
The performance of online algorithms is usually evaluated using competitive analysis. The \textit{competitive ratio} \cite{Borodin:1998} of an online algorithm is defined as the worst-case ratio between the cost of the solution obtained by the online algorithm versus that of an offline optimal solution obtained by knowing all input sequence in the future. 


\subsection{Alternative Current Electric Power Allocation}

There are several recent studies on demand response management for alternative current electric power sysetms with inelastic demands.
For a single-link case, demand response with inelastic demands has been studied as the complex-demand knapsack problem {\sc (CKP)} and its application to power demand allocation was highlighted by \cite{YC13CKP}. Let  $\theta$ be the maximum angle between any complex valued demands. \cite{YC13CKP} obtained a $\frac{1}{2}$-approximation for the case where $0 \le \theta \le \frac{\pi}{2}$.  \cite{woeginger2000does} (also \cite{YC13CKP}) proved that no fully polynomial-time approximation scheme (FPTAS) exist. Recently, \cite{CKM14, CKM15} provided a polynomial-time approximation scheme (PTAS), and a bi-criteria FPTAS (allowing constraint violation) for $\frac{\pi}{2} < \theta < \pi - \varepsilon$, which closes the approximation gap. An extension to radial network topology is considered in \cite{KCE2016OPF}, while the offline scheduling setting is considered in \cite{KKEC16CSPsss}.  

%% file: model.tex
\section{Model and Problem Formulation} \label{sec:model1}

The adopted DR model envisions a single LSE procuring the scheduling of customers' demands over a decision horizon $\cT\triangleq\{1,...,m\}$. The decision horizon $\cT$ is discretized into $m$ equal periods with a duration corresponding to the required time resolution granularity at which demand response management decisions are to be produced. To capture the volatile renewable energy sources, the available MG capacity is denoted by $C_t\in \RR$ at each time (slot) $t\in \cT$.

\subsection{Customer and Load Model} 

Consider a set of customers $\cN\triangleq\{1,...,n\}$ for a DR management scheme run by LSE. A customer $k\in \cN$ is associated with a complex-valued power demand $S_k \in \CC$ required for operating certain electric appliances at each time instant $t$ in a predetermined preferred schedulable interval $T_{k} = [t_1, t_2] \subseteq \cT$. Without loss of generality, we assume that the customers arrive in a sequential fashion from 1 to $n$. Also, it is assumed that each customer arriving at time $t$ declares $T_k$ at the {\it beginning} of $t$.

The customers' demands are categorized into two types according to their operation and energy consumption characteristics, \textit{elastic} (divisible) and \textit{inelastic} (indivisible). The demand of a customer possessing inelastic load can be either shed or fed completely over the specified time period. This models the electric appliances that can operate only under particular energy supply level (e.g., washing machine, vacuum cleaner). Different from the inelastic demands, an elastic load may be satisfied partially and adjusted to operate with different energy consumption levels (e.g., air conditioners, LED light bulbs).

Each customer is an independent decision maker. The response to the incentives posed by the DR aggregator is modeled by an \textit{utility function}. For simplicity, the utility function is summarized by an \textit{utility value} $u_k$ associated with a customer $k\in \cN$ that quantifies the extent of satisfaction obtained (or alternatively, the payment) by customer $k$ when own power demand is satisfied. In the case with inelastic demands, if $S_k$ is satisfied at each time $t \in T_{k}$, $u_{k}$ is the perceived utility for customer $k$, otherwise zero utility is perceived. As for a customer $k^{\prime} \in \cN$ with an elastic load, a portion $b \in [0,1]$ of the demand $S_{k^{\prime}}$ scheduled over $T_{k^{\prime}}$ imparts an utility of $b\cdot u_{k^{\prime}}$.

\subsection{Real-time Demand Response Management}

In the online setting, the set of customers $\cN$ is not known in advance, but is revealed progressively over time as customers continue to arrive. While DR participants arrive one at a time, LSE should determine, at each time instant, a scheduling decision on the demand arrived which once outputted cannot be undone or altered.

Note that deploying a centralized management system requires customers to submit their demands and utilities to LSE upon arrival as the inputs for optimization, which may compromise customer privacy. A distributed management system dilutes such concerns by allowing certain computations to be performed on a customer's side involving minimal exchanges of information. The algorithm proposed in this paper is sufficiently flexible to handle both centralized and distributed settings. For clarity of presentation, the centralized setting is primarily considered, where each customer declares his reactive and active power demand, preferred scheduling interval and his utility to LSE upon arrival.




\subsection{Optimization under Generation Capacity Constraint}
Now that the system model is established, the {\em complex-demand scheduling problem} \cite{KKEC16CSPsss} with a generation capacity constraint ({\sc CSP$_\textsc{C}$}) can be formulated by the following \textit{quadratically constrained integer programming} problem. 

\begin{align}
\textsc{(CSP$_\textsc{C}$)} \quad&  \displaystyle \max_{(x_k)_{k \in \cN}}  \sum_{k\in \cN } u_{k} x_{k} \nonumber\\
\text{subject to}  \quad  &  \displaystyle\bigg|\sum_{k\in \cN: t \in T_{k}} S_k x_{k}\bigg| \le C_t, \hspace{5pt} \forall t \in \cT \label{mc1}\\
& x_{k}\in\{0,1\} \quad\forall k\in \cN \label{mc3}
\end{align}
Here, $x_k$ is a binary decision variable that takes value $1$ if and only if the $k$-th customer's power demand $S_k$ is satisfied for all time slots $t \in T_k$. The {\sc CSP$_\textsc{C}$} problem aims at maximizing the overall net utility of customers arriving online without violating the apparent power generation $C_t$ over all time slots $t \in \cT$. Note that the time-varying capacity $(C_t)_{t \in \cT}$ is assumed to be known (or at least estimated) in advance. Otherwise, it is impossible to schedule a current customer's demand without violating the future capacity constraint.

Evidently, {\sc CSP$_\textsc{C}$} is {\sc NP-hard}, since the $0$-$1$ classical knapsack problem is a special case of {\sc CSP$_\textsc{C}$}. In fact, the presence of complex-demands in {\sc CSP$_\textsc{C}$} makes it an even harder problem (as shown to be strongly {\sc NP-hard} \cite{CKM15}). Before the proposed online algorithm is introduced, a measure for assessing the quality of the computed solution of {\sc CSP$_\textsc{C}$} problem is provided in the subsequent section.

\subsection{Competitive Ratio}

Let the inputs of {\sc CSP$_\textsc{C}$} at time $t$ of the arriving customers be $\sigma_t = \big( u_k, T_{k}, S_k\big)_{k \in \cN: t \in \cT}$. Recall that in an {\em online} algorithm, the decision at the current time $t$ only depends on the inputs available before or at $t$, namely, $(\sigma_{t'})_{t'\le t}$. Given input $\sigma = (\sigma_{t'})_{t'=1}^t$, let ${\mathbb E}[{\textsc{Alg}}[\sigma]]$ be the expected objective value (i.e., $\sum_{k\in \cN } u_{k} x_{k}$) by a randomized algorithm ${\textsc{Alg}}$, and ${\textsc{Opt}}(\sigma)$ be the objective value of an {\it offline optimal} solution (that knows all future inputs). In \textit{competitive algorithmic analysis}, the {\em competitive ratio} is a common performance metric, defined as the {\em worst-case} ratio between the expected objective value of the online algorithm ${\textsc{Alg}}$ and that of an offline optimal solution, namely,
\begin{equation}
{\tt CR}({\textsc{Alg}}) \triangleq \min_{\sigma} \frac{{\mathbb E}[{\textsc{Alg}}[\sigma]]}{{\textsc{Opt}}(\sigma)} \,.
\end{equation}
Similarly, an algorithm ${\textsc{Alg}}$ is called $c$-competitive, if ${\tt CR}({\textsc{Alg}}) = c$. The absence of any information concerning the future inputs limits the performance of an online algorithm severely as compared to that of offline algorithm possessing complete knowledge of all future inputs. In this context, comparing the performance of ${\textsc{Alg}}$ against such all-powerful benchmark is substantially difficult. In fact, it was shown in~\cite{Marchetti-Spaccamela1995} that if one makes no assumptions it is impossible to devise a non-trivial constant factor competitive online algorithm. This necessitates the need for introducing the assumptions listed hereunder that will be followed throughout this paper.

\begin{enumerate}

\item The largest demand of a customer is at most the smallest capacity over all time slots, namely,  $$\max_{k \in \cN} |S_k| \le C_{\min} \triangleq \min_{t \in \cT} C_t \, .$$
This is known in the literature \cite{Chakrabarti2002} as\textit{ the no bottleneck assumption} (NBA).

\item There exist positive $a^{\max}$, $a^{\min}$, $u^{\max}$, $u^{\min}$ and $T_{\max}$ known {\it apriori} such that for $\forall k \in \cN$
$$ a^{\min} \leq \frac{|S_k|}{u_k} \leq a^{\max},~ u^{\min} \leq u_k \leq u^{\max}\,, |T_k|\le T_{\max}.$$

\end{enumerate}

We remark that the NBA assumption naturally holds in power systems, since individual demands are typically much smaller than the generation capacity over all time slots.

%% file: alg-greedy.tex
\section{Competitive Online Algorithm}\label{sec:algs}

The presence of complex-valued power demands in {\sc CSP$_\textsc{C}$} problem bestows substantially challenging problem which, in general, is NP-hard to solve. This section presents an efficient randomized online algorithm (${\tt Online}$) in Algorithm~\ref{alg:onl} to compute solutions of {\sc CSP$_\textsc{C}$} problem that are close to the offline optimal solution, with a precise theoretical guarantee on their competitive ratio. ${\tt Online}$, which is extended from the algorithm presented in~\cite{Chakrabarti2002} for solving the Unsplittable Flow problem, as a subroutine invokes a primal-dual (${\tt PD}$) schema similar to that introduced in~\cite{TCS-024} for \textit{online fractional packing problem} ({\sc FPP}). The adapted ${\tt PD}$ schema is explained in Algorithm~\ref{alg:pds}.

Unlike {\sc FPP}, {\sc CSP$_\textsc{C}$} problem requires integral solutions. Also, {\sc CSP$_\textsc{C}$} has quadratic constraints, whereas {\sc FPP} has only linear ones. The basic idea is that ${\tt Online}$ relates the quadratically constrained {\sc CSP$_\textsc{C}$} to a linearly constrained packing problem by Lemma~\ref{lem:tb}. This allows us to use the framework of \cite{TCS-024} to obtain a close-to-optimal fractional solution $\widehat x$ (i.e., elastic demands). To convert the fractional solution $\widehat x$ to an integral solution $x$ (i.e., inelastic demands) without losing much in the quality of the solution, a rounding technique called {\it randomized rounding with correction} is utilized \cite{Chakrabarti2002}. For this to work, the demands are categorized based on a predetermined threshold into two sets, one set $\cI_L$ designated for those with {\it large} magnitude as compared to the capacity, while the remaining ones comprise the other set $\cI_S$. The primal-dual schema is then invoked on each set in parallel to obtain primal-dual fractional solutions $(\widehat x,\widehat y)$ and  $(\widetilde x,\widetilde y)$ for the small and large demands, respectively. Rounding the fractional solutions to an integral solution $x$ probabilistically concludes the execution.

We next show analytically that Algorithm ${\tt Online}$ is a competitive online algorithm for {\sc CSP$_\textsc{C}$} problem under certain assumptions. Define $\theta \triangleq \max_{k,k' \in \cN} |\arg(S_k) - \arg(S_{k'})|$ to be the maximum difference between the phase angles of any pair of customer power demands. Note that in practice $\theta < \frac{\pi}{2}$, due to regulations requiring electric equipment to conform with a maximum power factor. More precisely, $\theta$ is usually restricted to be in the range of $[0, 36^{\circ}]$ \cite{1339347}. Denote by $S_k^{\rm R} \triangleq \re(S_k)$ the\textit{ active power} demand of customer $k$, and by $S_k^{\rm I} \triangleq \im(S_k)$ the \textit{reactive power} demand. For clarity of presentation, this paper assumes (via a rotation) that $S_k^{\rm R} \ge 0$ and $S_k^{\rm I} \ge 0$ for $\forall t\in \cT, k\in \cN$.

\medskip

\begin{customthm}{1} \label{thm:online} Under the NBA assumption, algorithm ${\tt Online}$ produces a feasible solution to {\sc CSP$_\textsc{C}$}, with the following competitive ratio:
$$ {\tt CR}({\tt Online}) = 
\Omega\Bigg(\dfrac{\cos \tfrac{\theta}{2}}{\log(1+T_{\max} \frac{a^{\max}}{a^{\min}})}\Bigg).$$
\end{customthm}  

The proof of Theorem~\ref{thm:online} is deferred to the appendix.

\begin{algorithm}[htb!]
	\caption{${\tt Online}[k, u_{k}, T_k, S_k]$}
	\begin{algorithmic}[1]	
		\Statex \hspace{-15pt} {\bf Global Initialization:}
		\State $\widehat{x} \leftarrow {\bf 0}, \widetilde{x} \leftarrow {\bf 0},x \leftarrow {\bf 0}$; $\mathcal{I}_S \leftarrow \emptyset,\mathcal{I}_L  \leftarrow \emptyset$; \; $C^{\prime} \leftarrow (C_t)_{t\in \cT}$
		\State $\widehat y \leftarrow {\bf 0},\widetilde{y} \leftarrow {\bf 0}$; \; $a^{\max} \leftarrow \displaystyle \min\left\{1,\max_{j \in \cN} \Big\{\frac{|S_j|}{u_j}\Big\}\right\}$ 
		\State $a^{\min} \leftarrow \displaystyle \min\left\{1,\min_{j \in \cN} \Big\{\frac{|S_j|}{u_j} \, : |S_j| \neq 0\Big\}\right\}$; $s\leftarrow 0$, $l\leftarrow 0$
		\State $u^{\min} \leftarrow \displaystyle \min_{j \in \cN} \{u_j : u_j \neq 0\}$; $u^{\max} \leftarrow \displaystyle \max_{j \in \cN} \{u_j\}$; 
		\State $T^{\max}\leftarrow\max_{j \in \cN} \{|T_j|\}$; \; $\alpha\leftarrow 0.138$; \; $\delta\leftarrow 0.333$
		\medskip
		\State $r_S\leftarrow 2\log\Big(1 + \dfrac{(T^{\max}+1) a^{\max}}{a^{\min}}\Big)$  \label{alg:s1-1}
		\State $r_L\leftarrow 2\log\Big(1 + \dfrac{T^{\max} u^{\min}}{u^{\max}}\Big)$ \label{alg:s1-2}
		\State Choose $\tau\in\{0,1\}$ at random \label{alg:s0-}
		\Statex \hspace{-12pt}\textbf{Execution upon the $k$-th customer:}
		\vspace*{3pt}
		\If{$|S_k| \leq \delta \min_{t \in T_k} \{C_t \}$}
		\State $\mathcal{I}_S \leftarrow \mathcal{I}_S \cup \{k\}$ \Comment{{\em $\delta$-small demands}}
		\State  $s \leftarrow s + 1$; $\cT \leftarrow \cT \cup \{ |\cT| + 1\}$; \; $\widehat{T}_s \leftarrow T_k \cup \{ |\cT|\}$\label{alg:s0}
		\vspace*{5pt}
		\State $C_{|\cT|}\leftarrow u_k$; \; $a_{s,t} \leftarrow  \frac{|S_k|}{u_k} \, \forall t \in \widehat{T}_s $; \; $a_{s, |\cT|} \leftarrow 1$\label{alg:s1}
		\State $\widehat{x}_s,\widehat y \leftarrow {\tt PD}[s, ({C_t})_{t \in \cT}, (a_{i, t}, \widehat{T}_i)_{i \in \{1,...,s\}, j \in \cT}, \widehat{x}, \widehat y]$
		%
		\Else
		\State $\mathcal{I}_L \leftarrow \mathcal{I}_L \cup \{k\}$ \Comment{{\em $\delta$-large demands}}
		\State  $l \leftarrow l + 1$; \; $\widetilde{T}_l \leftarrow T_k$;\; $\widetilde{a}_{l,t} = \frac{1}{u_k} \, \forall t \in \widetilde{T}_l$
		\State $\widetilde{x}_l,\widetilde{y} \leftarrow {\tt PD}[l, (1)_{t \in \cT}, (\widetilde{a}_{i,t}, \widetilde{T}_i)_{i \in \{1,...,l\}, j \in \cT}, \widetilde{x}, \widetilde{y}]$
		%
		\EndIf
		
		\Statex \Comment{{\em Randomized rounding and correction}}
		\If{$k \in \mathcal{I}_L$}
		\State  $x_k \leftarrow \left\{
		\begin{array}{@{}rl@{}}
		1, & \mbox{with probability } \alpha\tau \frac{\widetilde{x}_l}{u_kr_L} \\
		0, & \mbox{with probability } 1-\tau\alpha \frac{\widetilde{x}_l}{u_kr_L} \\
		\end{array}
		\right.
		$
		\Else
		\State $x_k \leftarrow \left\{
		\begin{array}{@{}rl@{}}
		1, & \mbox{with probability } (1-\tau) \frac{\widehat{x}_s}{2u_kr_S} \\
		0, & \mbox{with probability } 1-(1-\tau)\frac{\widehat{x}_s}{2u_kr_S} \\
		\end{array}
		\right.
		$
		\EndIf
		
		\If{$\Big|\sum_{j\in\{1,...,k\}, t \in T_k} S_j x_j\Big| > C^{\prime}_{t} \text{ for some }  t \in \cT$}
		\State $x_k \leftarrow 0$
		\Else \If{$x_k = 1$}
		\State $C^{\prime}_{t} \leftarrow C^{\prime}_{t} - |S_k|$ $\forall t \in T_k$
		\EndIf
		\EndIf
		
		\State \Return $x_k$
	\end{algorithmic}
	\label{alg:onl} 
\end{algorithm}

\begin{algorithm}[htb!]
	\caption{${\tt PD}[k, (\bar C_t)_{t \in \cT}, (a_{i, t}, \bar T_i)_{i \in \{1,...,k\}, j \in \cT}, x, y]$}
	\begin{algorithmic}[1]
		\State $\bar a^{\max} \leftarrow \displaystyle \max_{j\in\{1,...,k\}, t \in \bar T_j} \{ a_{j, t}\} \, $; $\bar T^{\max}\leftarrow\max_{1\leq j\leq k} \{|\bar T_j|\}$
		\While{$\sum_{t \in \bar T_k} y_t a_{k, t} < 1$}
		\State Increase $x_k$ continuously 
		\For{$t \in \bar T_k$}
		\medskip
		\State $b \leftarrow e^{(2\bar C_{t})^{-1}\sum_{j\in \{1,...,k\}, t \in\bar T_j} a_{j,t} x_j}$
		\State $y_{t} \leftarrow \max \Big\{ y_{t}, \dfrac{b-1}{\bar T^{\max} \bar a^{\max}}\Big\}$
		\EndFor
		
		\EndWhile
		\State \Return $x_k$, $y$
	\end{algorithmic}
	\label{alg:pds} 
\end{algorithm}


%% file: voltage.tex
\section{Optimization under Voltage Constraint}

This section presents a variant of the complex-demand scheduling problem where, instead of the capacity constraint, the nodal voltages in the distribution network are restricted to remain within the nominal range at all time steps. To introduce the voltage constraint into the DR optimization problem a model of the distribution network is established below.  
 

Consider a path topology $\cG=(\cV,\cE)$ that represents a distribution system feeder, where each customer $k$ is located at a given node except the root. For simplicity, assume $\cV=\{0,1,\ldots,d\}$ and $\cE=\{(0,1),(1,2),\ldots,(d-1,d)\}$. The root (denoted by node $0$) is either a substation or a generator which powers the entire system. For each node $i\in\cV\backslash\{0\}$, there is a set of customers attached to $i$, denoted by $\cM_i$.
For node $i \in \cV$, denote its voltage by $V_i \in \CC$. For each edge $e=(i,i+1) \in \cE$, denote its current from $i$ to $i+1$ by $I_{i,i+1}$, its impedance by $z_{i,i+1} \in \CC$, and with a slight abuse of notation, its transmitted power by $\widehat S_{i,i+1}$.
A power flow in a steady state is described by a set of power flow equations. In radial networks (which include paths) the Branch Flow Model (BFM) proposed by \cite{baran1989placement} can be used to model them.



Let $v_i \triangleq |V_i|^2$ and $\ell_{i,i+1} \triangleq |I_{i,i+1}|^2$ be the magnitude square of voltage at node $i\in \cV$ and current at edge $(i,i+1)\in \cE$, respectively. 
The BFM is given by the following for all $(i,i+1) \in \cE$
\begin{align}
\ell_{i,i+1}& =  \frac{|\widehat S_{i,i+1}|^2}{v_i},   \\
v_{i+1}& =  v_i + |z_{i,i+1}|^2 \ell_{i,i+1} - 2 \re(z_{i,i+1}^\ast  \widehat S_{i,i+1}),  \label{eq:vj}\\
\widehat S_{i,i+1}& = \widehat S_{i+1,i+2} + \sum_{k \in \cM_{i+1}} S_kx_k + z_{i,i+1}\ell_{i,i+1},  \label{eq:bf3}
\end{align}
where it is assumed $\widehat S_{d,d+1}=0$. Here, $x_k$ again is a binary decision variable that takes value $1$ if and only if the $k$-th customer's power demand $S_k$ is satisfied. 
At any time step, the operating constraint requires voltage $v_i$ to be at least a given number $v_{\min} \in \RR$, that is 
\begin{equation}
v_i \ge v_{\min}, \, \, \,\forall i=1,\ldots,d \,. \label{eq:cv}
\end{equation}
In MGs, the size of the grid is typically small and most power demand can be attributed to customers’ demands, and hence, the effect of transmission loss is negligible. As a consequence, one can approximate the optimal power flow model by dropping the terms associated with the transmission power loss (i.e., $|z_{i,i+1}|^2 \ell_{i,i+1}, z_{i,i+1}\ell_{i,i+1}$), and then derive a feasible solution without explicitly considering the transmission power loss \cite{KCE2016OPF}. Rewriting Eqn.~\raf{eq:bf3} by recursively substituting $\widehat S_{i+1,i+2}$ and dropping $z_{i,i+1}  \ell_{i,i+1}$ gives $\widehat S_{i,i+1} = \sum_{j=i+1}^d\sum_{k\in \cM_j} S_k x_k\, \, \,\forall (i,i+1) \in \cE \label{eq:se}$. Also, rewriting Eqn.~\raf{eq:vj} recursively by substituting $v_i$ and dropping all loss terms yields $v_i = v_0 -2\sum_{ j=0 }^{i} \re(z_{j,j+1}^\ast  \widehat S_{j,j+1}),\, \, \,\forall i=1,\ldots,d$.
 
For $k\in\cN$ denote by $\ell(k)$ the index $\ell \in \cV$ such that $k\in\cM_\ell$. For each $i=1,\ldots d$, Cons.~\raf{eq:cv} can be rewritten as
{\small
\begin{equation}
\sum_{ j=0 }^{i} \re(z_{j,j+1}^\ast  \widehat S_{j,j+1})   \le \tfrac{1}{2}(v_0 - v_{\min}) \label{eq:cv-1}
\end{equation}	
}
{\small
\begin{align}
\Rightarrow & \sum_{ j=0}^i \sum_{\ell=j+1}^d\sum_{k\in \cM_\ell} (z^{\rm R}_{j,j+1} s_k^{\rm R}+ z^{\rm I}_{j,j+1} s_k^{\rm I} ) x_k  \le \tfrac{1}{2}(v_0 - v_{\min})  \\
\Rightarrow  & \sum_{k \in \cN} \Big( \sum_{ j=0 }^{h} (z^{\rm R}_{j,j+1} s_k^{\rm R}+ z^{\rm I}_{j,j+1} s_k^{\rm I}) \Big) x_k  \le \tfrac{1}{2}(v_0 - v_{\min})\,, \label{eq:cv-3}
\end{align}}
\noindent where $h \triangleq \min\{i,\ell(k)-1\}$ and the last statement follows from exchanging the summation operators. 

\medskip

\begin{assumption}\label{as:v}
	We assume throughout this paper that $ z^{\rm R}_{i,i+1} S_k^{\rm R}+ z^{\rm I}_{i,i+1} S_k^{\rm I} \ge 0$ for all customers $k$ and edges $(i,i+1)$. 
\end{assumption}
\medskip

Observe that by Assumption~\ref{as:v} it is sufficient to consider Cons.~(\ref{eq:cv-3}) only at the edge $(d-1,d)\in\cE$ furthest from the root, because the left hand side of the constraint at that edge is the largest among all other edges. The complex-demand scheduling problem under voltage constraint ({\sc CSP$_\textsc{V}$}) is embodied by the following integer programming problem.
{\small
\begin{align}
\textsc{(CSP$_\textsc{V}$)} &  \displaystyle \max_{(x_k)_{k \in \cN}}  \sum_{k\in \cN } u_{k} x_{k} \nonumber\\
\text{s.t} \,\,  &   \sum_{\substack{k \in \cN\\ t \in T_k}} \Big( \sum_{ j=0  }^{\ell(k)-1} z^{\rm R}_{j,j+1} S_k^{\rm R}+ z^{\rm I}_{j,j+1} S_k^{\rm I} \Big) x_k \le  \widehat v, \,\,\forall t \in \cT\\
& x_{k}\in\{0,1\} \quad\forall k\in \cN \, ,\label{vc2}
\end{align}}
where $\widehat v \triangleq \tfrac{1}{2}(v_0 - v_{\min})$.

\begin{figure}[!htb] 
	\begin{center}
		\includegraphics[scale=.43]{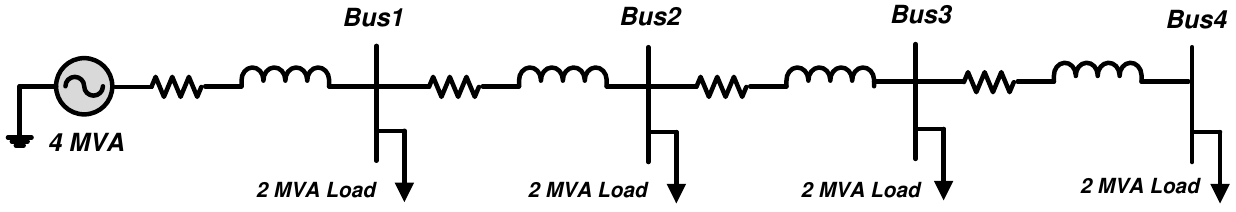}
	\end{center} 
	\caption{A 4-bus from Canadian benchmark distribution system.}
	\label{fig:system}
\end{figure}

A similar algorithm of Algorithm~\ref{alg:onl} can be devised to solve {\sc CSP$_\textsc{V}$}. It is identical to ${\tt Online}$ except that the customer power demands $S_k$ are replaced by $z^{\rm R}_{j,j+1} S_k^{\rm R}+ z^{\rm I}_{j,j+1} S_k^{\rm I}$ for $\forall k \in \cN$ and time-varying capacities $C_t$ are replaced by $\widehat v$ over all the time slots $t \in \cT$. The consecutive section 
applies the modified algorithm to one of the feeders of the Canadian benchmark system (see Fig.~\ref{fig:system}).

%% file: sims.tex
\begin{figure*}
	\centering
	\begin{minipage}{.49\textwidth}
		\centering
		\includegraphics[scale=.47]{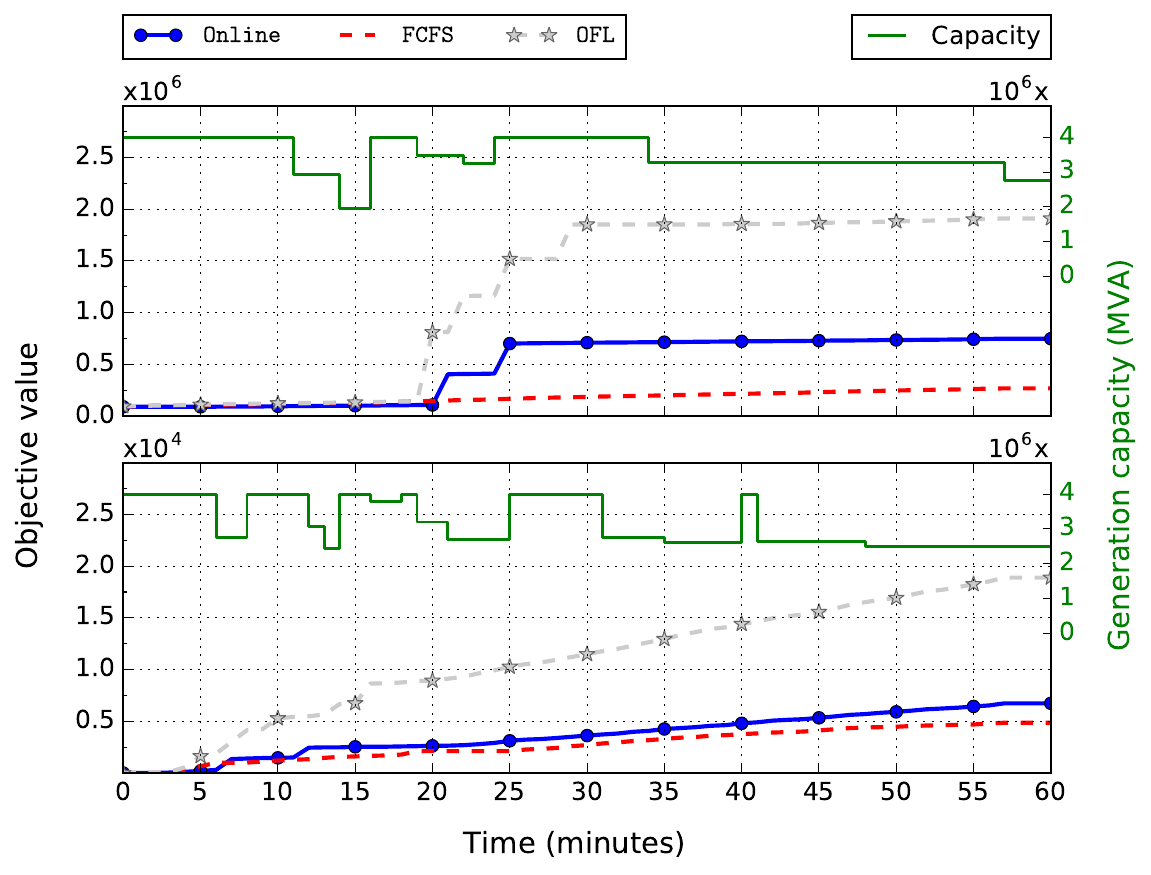} 
		\captionof{figure}{The objective values of ${\tt Online}$, ${\tt FCFS}$, and ${\tt OFL}$ when applied to {\sc CSP$_\textsc{C}$} for the case studies with quadratic and random customer utilities on top and bottom respectively.}
		\label{fig:obj}
	\end{minipage}%
	\hspace*{5pt}
	\begin{minipage}{.49\textwidth}
		\centering
		\includegraphics[scale=.475]{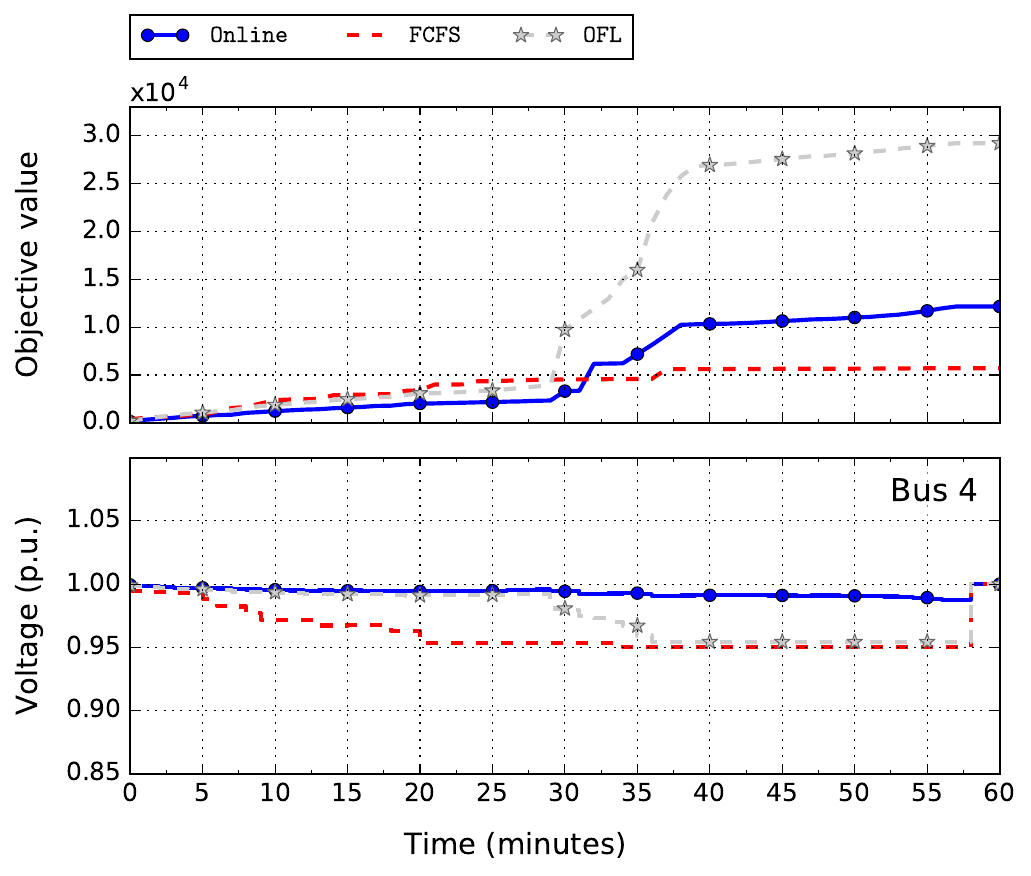} 
		\captionof{figure}{The objective values and voltage profile on bus $4$ of ${\tt Online}$, ${\tt FCFS}$, and ${\tt OFL}$ when applied to {\sc CSP$_\textsc{V}$} problem considering random customer utilities.}
		\label{fig:objv}
	\end{minipage}
	\vspace*{-15pt}
\end{figure*}

\section{Empirical Evaluation}\label{sec:results}
To complement the analytic result in Theorem~\ref{thm:online}, this section evaluates the proposed algorithm ${\tt Online}$ empirically under extensive scenarios. The objective value obtained by ${\tt Online}$ when applied to {\sc CSP$_\textsc{C}$} and {\sc CSP$_\textsc{V}$} is compared against that of offline optimal solution by numerical solver {\sc Cplex}. The results are also compared with a benchmark algorithm, ${\tt FCFS}$ (first-come-first-serve), which greedily schedules the loads in the arrival order until the corresponding operating constraint is violated (i.e. apparent power constraint or voltage constraint).

\subsection{Simulation Setup}
An MG with an overall generation capacity of $4$MVA is simulated with over $2000$ customers arriving in a sequential manner. Each customer has a power demand (including both active and reactive power) defined over a certain time interval and a utility that is generated according to a probability preference model. The output solution of {\sc Cplex} optimizer is denoted by ${\tt OFL}$. Customer arrival follows an uniformly random probability distribution and demand durations are picked at random, from a uniform distribution on the interval $\{1, \hdots , \max_{k \in \cN} |T_k|\}$ where $|T_k| \leq |\cT|$ for $\forall k \in \cN$. The time-varying generation capacity follows a \textit{Bernoulli} process.
 
For the {\sc CSP$_\textsc{V}$} problem the candidate algorithms are applied to the Canadian benchmark distribution system depicted in Fig~\ref{fig:system}. The power system analysis package {\sc Pscad} was utilized to simulate the feeder which is rated at $8.7$MVA, $400$A and $12.47$KV. Each feeder section is a $700$MCM Cu XLPE cable with $z = 0.1529 + J0.1406$ $\Omega$/km and each node consists of a $2$MVA total load.


The following are settings for the case studies in this paper.
\begin{itemize}
	\item[a)] \textit{Quadratic utility}: The utility of each customer is a function of the power demand in the form of $$u_k(|S_k|) = a |S_k|^2 + b|S_k| + c \, ,$$ where $a, b, c \geq 0$ are predetermined constants.
	\item[b)] \textit{Random utility}: The utility of each customer is independent of the power demand and is generated randomly from $[0, |S_{\max}(k)|]$, where $|S_{\max}(k)|$ depends on the customer type: if customer $k$ is a \textit{commercial} customer then $|S_{\max}(k) = 1\text{MVA}|$, otherwise if \textit{residential} $|S_{\max}(k) = 20\text{KVA}|$.
\end{itemize}

\subsection{Results}
Figs.~\ref{fig:obj} and~\ref{fig:objv} illustrate the empirical results for two case studies with quadratic and random utilities. As can be observed from both figures, ${\tt Online}$ approaches ${\tt OFL}$ in most cases, whereas ${\tt FCFS}$ drifts far away from ${\tt OFL}$ in some cases. The results illustrated in Fig.~\ref{fig:objv} show that ${\tt Online}$ obtains higher objective value as compared to ${\tt FCFS}$ while maintaining appropriate voltage level of the end of the feeder (i.e. is within IEEE standard 1547 limits).







%% file: concl.tex
\section{Conclusion}

We presented a competitive randomized online algorithm for deciding whether a sequence of inelastic demands can be allocated for the requested intervals, subject to the total satisfiable apparent power within a time-varying capacity constraint. We also applied to an alternative setting with nodal voltage constraint, using a variant of the online algorithm.  In future work, we will consider an extension to radial network topology \cite{KCE2016OPF}. Finally, we note that our online algorithm can be adapted to a distributed setting, such that the customers decide the scheduling decisions without submitting full preferences. A detailed adaption will be provided in future work.

%% file: append.tex
\appendix 

The appendix gives an overview of the preliminaries of primal-dual schema \cite{TCS-024} followed by the proof of Theorem \ref{thm:online}.


\subsection{Duality and Linear Relaxation}

Relax the integrality constraints~\raf{mc3} $(x_{k})_{ k\in \cN} \in\{0,1\}^{|\cN|}$ in {\sc CSP$_\textsc{C}$} problem and substitute customers' complex-valued power demands by their real-valued magnitude:
{\small
\begin{align}
\textsc{(CSP$_\textsc{R}$)} \quad&  \displaystyle \max_{({x}_k)_{k \in \cN}}  \sum_{k\in \cN } u_k x_k \nonumber\\
\text{subject to}  \quad  &  \displaystyle\sum_{k\in \cN: t \in T_{k}} |S_k| x_{k} \le C_t, \hspace{5pt} \forall t \in \cT \label{pmc11}\\
& 0\leq {x}_{k}\leq 1 \quad\forall k\in \cN \label{pmc31} \, .
\end{align}}
Note that any feasible integral solution $x$ for \textsc{(CSP$_\textsc{R}$)} is also feasible for {\sc CSP$_\textsc{C}$}, as $|\sum_{k\in \cN: t \in T_{k}} S_k x_{k}|\le\sum_{k\in \cN: t \in T_{k}} |S_k| x_{k}\le C_t$ for all $t\in\cT$.
 
Define $\widehat{x}_k \triangleq u_k x_k$.  
We will use the following generic linear programming problem (which we call the {\it dual LP} for convenience). The coefficients $a_{k,t}$, the sets $\bar\cN$, $\bar T_k$ and $\bar\cT$, and the right-hand sides $\bar C_t$ will be set later depending on whether we are dealing with large or small demands.
{\small
\begin{align}
\textsc{(D)} \quad&  \displaystyle \max_{(\widehat{x}_k)_{k \in \cN}}  \sum_{k\in \bar\cN } \widehat{x}_k \nonumber\\
\text{subject to}  \quad  &  \displaystyle\sum_{k\in \cN: t \in \bar T_{k}} a_{k,t} \widehat{x}_{k} \le \bar C_t, \hspace{5pt} \forall t \in \bar\cT \label{pmc1}\\
& \widehat{x}_{k}\geq 0 \quad\forall k\in \bar\cN \label{pmc3} \, .
\end{align}}
Next, formulate the corresponding {\it primal} minimization linear programming problem, dual to ({\sc D}).

\begin{align}
\textsc{(P)} \quad&  \displaystyle \min_{(y_t)_{t \in \bar\cT}}  \sum_{t\in \bar\cT } \bar C_{t} y_{t} \nonumber \\
\text{subject to}  \quad  &  \displaystyle\sum_{t \in \bar T_{k}} a_{k,t} y_{t} \geq 1, \hspace{5pt} \forall k \in \bar\cN \label{dmc1}\\
& y_{t}\geq 0 \quad\forall t\in\bar\cT \label{dmc3} \, .
\end{align}

The primal-dual pair ({\sc P}) and ({\sc D}) are broadly referred to as \textit{general fractional covering and packing problems} respectively. Observe that the primal linear program ({\sc P}) comprises $|\bar\cT|$ variables which correspond to the constrains of the dual  ({\sc D}) and $|\bar\cN|$ constraints corresponding to the dual variables. A well-known fundamental lemma which establishes the correlation between primal and dual feasible solutions is stated below. 
\vspace*{5pt}

\begin{customlem}{1} \textit{(Weak duality)}
	\label{weak}
	Let $\widehat{x}$ and $y$ be feasible solutions satisfying the constrains of the dual and the primal problems ({\sc D}) and ({\sc P}) accordingly, then
	$$\sum_{k\in\bar\cN} \widehat{x}_k \leq \sum_{t\in\bar\cT} {\bar C}_t y_t\,.$$
\end{customlem}

\subsection{Primal-Dual Schema}

In this section, we describe the primal-dual schema of \cite{TCS-024} for providing a competitive solution for the online fractional packing problem (\textsc{D}).

Recall that in the online setting studied here the constraints matrix defining the feasibility region of ({\sc D}) is revealed to the algorithm column by column. In other words, the knowledge about the capacity profile and time interval is available to ${\tt Online}$ prior to execution, whereas customer utilities and demands are given one at a time. The ${\tt Online}$ algorithm at each round outputs a load scheduling decision on only the current demand, which once made is irrevocable. 

Consider the $k$-th round where customer $k \in \bar\cN$ arrives introducing a new variable $\widehat{x}_k$ along with its respective constraints. ${\tt Online}$ first obtains a fractional solution $\widehat{x}_k$ by solving the primal-dual pair ({\sc P}) and ({\sc D}). The algorithm attains this by increasing the value of the new variable $\widehat{x}_k$ continuously and updating the primal variables $(y_t)_{t \in \bar\cT}$ such that
\begin{equation}\label{updrl}
y_{t} = \max \Big\{ y_{t}, \dfrac{e^{\frac{1}{2 \bar C_{t}}\sum_{j\in\{1, ...,k\}, t \in \bar T_j} a_{j,t} \widehat{x}_j}-1}{\bar T^{\max} \bar a^{\max}}\Big\} \, ,
\end{equation}
as long as the corresponding primal constrains are infeasible. Let $\mathit P(k)$ and $\mathit D(k)$ be the objective values of the primal and dual problems respectively, obtained up to the $k$-th round. In \cite{TCS-024}, the following three claims were proved to hold for the aforementioned primal-dual schema.

\begin{enumerate}
	
	\item[({\sf A1})] In each iteration $k$, $\mathit P(k) \leq \mathit D(k)$.
	\item[({\sf A2})] The scheme produces a feasible primal solution.
	\item[({\sf A3})] The following holds for the generated dual solution,  for all $t\in \bar\cT$: 
	$$ \sum_{j\in\{1,...,k\}, t\in \bar T_j} a_{j, t}\widehat{x}_j \leq \bar C_t\cdot \widehat r$$
	where $\widehat r\triangleq2\log\Big(1 + \frac{\bar T_{\max}\bar a^{\max}}{\bar a^{\min}}\Big)$.

\end{enumerate}

\noindent For completeness, we give here the proofs of ({\sf A1}), ({\sf A2}), ({\sf A3}), adopted to our setting.	

\medskip

\textbf{\textit{Proof of}} ({\sf A1}): 
Given that initially, $\mathit P(0) = \mathit D(0) =0$, consider the $k$-th round when the dual variable $\widehat{x}_k$ is being increased iteratively. To prove the claim it suffices to show that $\frac{\partial \mathit P(k) }{\partial \widehat{x}_k} \leq \frac{\partial \mathit D(k)}{\partial \widehat{x}_k}$. It follows from the way we defined the update rule (Eqn.~\raf{updrl}) of primal variables that 
$\frac{\partial y_t}{\partial \widehat{x}_k}~\leq~\frac{1}{\bar T_{\max} \bar a^{\max}} \frac{\partial (e^{\frac{1}{2 \bar C_{t}}\sum_{j\in\{1, ...,k\}, t \in \bar T_j} a_{j,t} \widehat{x}_j}-1)}{\partial \widehat{x}_k}$ for all $t \in \bar T_k$, and thus,
{\small
\begin{align}
& \frac{\partial \mathit P(k) }{\partial \widehat{x}_k} =  \sum_{t \in \bar\cT} {\bar C}_t \frac{\partial y_t }{\partial \widehat{x}_k} \nonumber \\
  \leq & \sum_{t \in \bar T_k} \frac{{\bar C}_t a_{k,t}}{2\bar C_t}\frac{ e^{\frac{1}{2 \bar C_{t}} \underset{j\in\{1, ...,k\}, t \in \bar T_j}{\sum} a_{j,t} \widehat{x}_j} }{\bar T_{\max} \bar a^{\max}} \nonumber \\
  = & \frac{1}{2} \sum_{t \in \bar T_k} a_{k,t} \Big[\frac{\Big(e^{\frac{1}{2 \bar C_{t}} \underset{j\in\{1, ...,k\}, t \in\bar T_j}{\sum} a_{j,t} \widehat{x}_j}  -1\Big)}{\bar T_{\max} \bar a^{\max}} +  \frac{1}{\bar T_{\max}\bar a^{\max}} \Big] \,. \label{eq1}
\end{align}}
Since by~\raf{updrl} $y_t \geq \dfrac{e^{\frac{1}{2 \bar C_{t}}\sum_{j\in\{1, ...,k\}, t \in\bar T_j} a_{j,t} \widehat{x}_j}-1}{\bar T_{\max} \bar a^{\max}}$, rewriting Eqn.~\raf{eq1} gives
{\small
\begin{eqnarray}
\frac{\partial \mathit P(k) }{\partial \widehat{x}_k}& \leq & \frac{1}{2} \sum_{t \in \bar T_k} a_{k,t} \Big[y_t +  \frac{1}{\bar T_{\max}\bar a^{\max}} \Big] \nonumber \\ 
& = &  \frac{1}{2}\Big(\sum_{t \in\bar T_k} a_{k,t}y_t + \frac{1}{\bar T_{\max}}\sum_{t \in\bar T_k}\frac{a_{k,t}}{\bar a^{\max}}\Big) \nonumber \\
\label{eq2}
& \leq & \frac{1}{2}\Big(\sum_{t \in\bar T_k}a_{k,t}y_t + 1\Big) \,,
\end{eqnarray}
}
\noindent as $|\bar T_k|\le \bar T^{\max}$.
Since the covering constraint~\raf{dmc1} remains infeasible while $\widehat{x}_k$ is being iteratively increased, it follows from Eqn.~\raf{eq2} that

\begin{eqnarray}
\frac{\partial \mathit P(k) }{\partial \widehat{x}_k}& \leq & \frac{1}{2}\Big(1 + 1\Big) =  1 = \frac{\partial \mathit D(k) }{\partial \widehat{x}_k}\,.
\end{eqnarray}

\medskip

\noindent\textbf{\textit{Proof of}} ({\sf A2}): Consider the $k$-th round when the corresponding primal variables $y_t$ are being increased until the associated primal constraints become feasible. The feasible constrains remain satisfied in the subsequent rounds since none of these variables decrease.

\medskip

\noindent\textbf{\textit{Proof of}} ({\sf A3}): Note that during the execution of this primal-dual schema a primal variable $y_t$, for $t\in\bar T_k$, is upper bounded by $\frac{1}{\bar a^{\min}}$ (i.e., $y_t \leq \frac{1}{\bar a^{\min}}$) in the $k$-th round, since otherwise the corresponding constraint $a_{k,t}y_t \geq 1$ with $a_{k,t} >0$ would already be satisfied. Therefore,
\begin{eqnarray}
  \dfrac{e^{\frac{1}{2 C_{t}}\sum_{j\in\{1,...,k\}, t \in \bar T_j}^k a_{j,t} \widehat{x}_j}-1}{\bar T^{\max}\bar a^{\max}} \leq y_t \leq \frac{1}{\bar a^{\min}}\,. \label{eq3}
\end{eqnarray}
Simplifying Eqn.~\raf{eq3} yields
 \begin{eqnarray}
 \sum_{j\in\{1,...,k\},t \in\bar T_j} a_{j,t}\widehat{x}_j \leq 2\log\Big(1 + \frac{\bar T^{\max}\bar a^{\max}}{\bar a^{\min}}\Big) \cdot \bar C_t \,.\label{eq4} \nonumber
 \end{eqnarray}
Scale down each dual variable $\widehat{x}_k$ by a factor of $\widehat r=2\log\Big(1 + \frac{\bar T^{\max}\bar a^{\max}}{\bar a^{\min}}\Big) $ to construct a dual feasible solution. Then, by claim ({\sf A1}), we obtain
\begin{equation}
\label{eq5}
\sum_{k \in \bar\cN} \frac{1}{\widehat r}\widehat{x}_k\geq \frac{1}{\widehat r}\sum_{t\in \bar\cT} \bar C_t y_t\, .
\end{equation}
Let $\Opt_D$ be the objective value of an optimal offline solution to the dual problem ({\sc D}). Lemma~\ref{weak} together with claim ({\sf A2}) imply that 
\begin{equation}
\label{eq6}
\sum_{t\in \bar\cT} \bar C_t y_t \geq \Opt_D\,,
\end{equation}
and thus it follows from Eqns.~\raf{eq5} and~\raf{eq6} that
\begin{equation}
\sum_{k \in \bar\cN} \frac{\widehat{x}_k}{\widehat r}\geq \frac{\Opt_D}{\widehat r}\, .
\end{equation}
Hence, this proves that the primal-dual schema above devises an $\widehat r$-competitive solution for the dual problem ({\sc D}).

\subsection{Proof of Theorem \ref{thm:online}}

\begin{proof}
The proposed algorithm ${\tt Online}$ when applied to {\sc CSP$_\textsc{C}$} problem employs the above primal-dual schema, after partitioning demands into two sets $\mathcal{I}_S$ and $\mathcal{I}_L$. Following \cite{Chakrabarti2002}, given a constant $\delta \in [0,\frac{1}{2})$, we define a demand $S_k$ to be \textbf{$\delta$-small} if $|S_k| \le \delta \min_{t \in T_k} \{C_t \}$, otherwise define it to be \textbf{$\delta$-large}. The former set $\mathcal{I}_S$ is comprised of only the $\delta$-small demands, and the latter set $\mathcal{I}_L$ contains the rest of the demands. The primal-dual schema is applied on each set in parallel. The algorithm then rounds the obtained solutions randomly under NBA and corrects the rounded integral solutions whenever necessary. Next, we analyze these three steps in detail. 

We denote by ~\textsc{(CSP$_\textsc{R}$[S])} and~\textsc{(CSP$_\textsc{R}$[L])} the instantiations of  
problem~\textsc{(CSP$_\textsc{R}$)}, when the set of demands $\cN$ is replaced by $\cI_S$ and $\cI_L$, respectively.
Let $\Opt$, $\Opt^*$, $\Opt_S^*$ and $\Opt_L^*$ be the objective values of an optimal (offline) solution to problems \textsc{(CSP$_\textsc{C}$)}, \textsc{(CSP$_\textsc{R}$)}, \textsc{(CSP$_\textsc{R}$[S])} and~\textsc{(CSP$_\textsc{R}$[L])}, respectively.

\medskip

\noindent\textbf{$\delta$-Small Demands:} To obtain a feasible approximate fractional solution $(\widehat{x}_k)_{k \in \mathcal{I}_S}$ for $\delta$-small demands ${\tt Online}$ applies the aforementioned primal-dual schema on an instance of problem ({\sc D}) with $\delta$-small demands only with the following parameters: $\bar\cN=\cI_S$, $a_{k,t} = \frac{|S_k|}{u_k}$ for $\forall k \in \cI_S$, $t\in T_k$ and $\bar C_t=C_t$ for all $t\in\cT$. To account for the constraint $x_k\le 1$, we imagine a dummy time slot $t$ with capacity $1$ (hence, the assignments $\bar T_s\leftarrow T_k\cup\{|\cT|\}$, $C_{|\cT|}\leftarrow u_k$ and $a_{s,|\cT|}\leftarrow 1$ in steps~\ref{alg:s0}-\ref{alg:s1} of Algorithm~\ref{alg:onl}); this ensures the normalized fractional decision variable $\widehat{x}_k$ is upper bounded by $u_k$. 
Thus, $\Opt_D={\Opt_S^*}$.

Let $\widehat x$ be the solution returned by Algorithm~\ref{alg:pds} on $\delta$-small demands, and set $x^s=(\frac{\widehat x_k}{u_k})_{k\in\cI_S}$. 
Then by ({\sf A1}), ({\sf A2}) and ({\sf A3}) above, $\sum_{k\in\cI_S}u_kx_k^s=\sum_{k\in\cI_S}\widehat x_k^s\ge\frac{\Opt_D}{r_S}= \frac{\Opt_S^*}{r_S}$, where $r_S$ is given in step~\ref{alg:s1-1}.

\vspace*{5pt}
\noindent\textbf{$\delta$-Large Demands:} To obtain a feasible approximate fractional solution $(\widetilde{x}_k)_{k \in \mathcal{I}_L}$ for $\delta$-large demands ${\tt Online}$ applies the aforementioned primal-dual schema on an instance of problem ({\sc D}) with $\delta$-large demands only with the following parameters: $\bar\cN=\cI_L$, $a_{k,t} = \frac{1}{u_k}$ for $\forall k \in \cI_L$, $t\in T_k$ and $\bar C_t=1$ for all $t\in\cT$.  
In other words, ${\tt Online}$ invokes the primal-dual approach to approximately solve \textit{fractional weighted maximum independent set} ({\sc FWMIS}) problem on the set $\mathcal{I}_L$. Define $\widetilde{x}_i \triangleq u_i x_i \, \forall i \in \mathcal{I}_L$, then {\sc FWMIS} problem is captured by the following linear programming problem.
{\small
\begin{align}
\textsc{(FWMIS)} \quad&  \displaystyle \max_{(\widetilde{x}_k)_{k \in \mathcal{I}_L}}  \sum_{i\in \mathcal{I}_L } \widetilde{x}_k \nonumber\\
\text{subject to}  \quad  &  \displaystyle\sum_{k\in \mathcal{I}_L: t \in T_{k}} \frac{\widetilde{x}_{k} }{u_k}\le 1, \hspace{5pt} \forall t \in \cT \label{pmc2} \nonumber\\
& \widetilde{x}_{k}\geq 0 \quad\forall k\in \mathcal{I}_L \nonumber \, . 
\end{align}}
Denote by $\Opt_F$ the objective value of an optimal solution of ({\sc FWMIS}). Then $\Opt_D=\Opt_F\ge\frac{\delta^2}{2}\Opt^*_L$ by part ({\sf B2}) of Lemma \ref{lem:l1} below.

Let $\widetilde x$ be the solution returned by Algorithm~\ref{alg:pds} on $\delta$-large demands, and set $x^\ell=(\frac{\widetilde x_k}{u_k})_{k\in\cI_L}$. 
Then by ({\sf A1}), ({\sf A2}) and ({\sf A3}) above, $\sum_{k\in\cI_L}u_kx_k^\ell=\sum_{k\in\cI_L}\widetilde x_k^\ell\ge\frac{\Opt_D}{r_L}= \frac{\delta^2\Opt_L^*}{2r_L}$, where $r_S$ is given in step~\ref{alg:s1-2}.

\medskip

\noindent\textbf{Randomized Rounding and Correction:} 
Initially, ${\tt Online}$ (step~\ref{alg:s0-}) chooses $\tau\in\{0,1\}$ at random to determine which of the fractional solutions, either among $\delta$-large demands or $\delta$-small demands, will be used for rounding.
${\tt Online}$ lastly randomly rounds the fractional solutions of the chosen set (either $\delta$-large or $\delta$-small demands), after scaling them by a factor of $\frac{\alpha}{u_k}$ ($\alpha \leq 1$), to obtain an integral solution $(x_i)_{i \in \cN} \in \{0,1\}^n$ to problem {\sc CSP$_\textsc{C}$}.  This may still result in infeasible solution, which calls for a correction step where a manual feasibility check should be invoked. In this step, the demand arrived in $k$-th round is scheduled (i.e, $x_k = 1$) over the time interval $T_k$ if scheduling it does not violate any capacity constraint. In the case if the demand is accepted, the corresponding capacities are reduced by the magnitude of the demand to maintain feasibility.

Denote by $Z^{{\tt ONL}}$ the expected utility of the output solution of ${\tt Online}$ for the {\sc CSP$_\textsc{C}$} problem. In~\cite{Chakrabarti2002} it was shown that the rounding steps for $\delta$-small demands in ${\tt Online}$ result in an integer solution (among $\cI_S$) with an expected total utility of at least ${\alpha}\Big(1-\sum_{i \, \geq \, 0}^{}\Big(\frac{(\frac{1}{2} - \delta - \alpha)}{\alpha}\Big)^{2^i}\Big)\frac{\Opt^*_S}{r_S}\ge 0.0035\frac{\Opt^*_S}{r_S}$ for $\alpha=0.138$ and $\delta=0.333$. On the other hand, for $\delta$-large demands, Lemma~\ref{lem:l3} below shows that the expected utility of rounded integer solution ${\tt Online}$ obtains for $\delta$-large demands is at least $\frac{1}{4}\Opt_F$ which is at least $\frac{\delta^2}{8r_L}\Opt_L^*=0.0139\frac{\Opt_L^*}{r_L}$. Since ${\tt Online}$ chooses among the $\delta$-large and $\delta$-small demands with equal probability, 
{\small
\begin{align}
Z^{{\tt ONL}}  \geq & \frac{1}{2}(0.0035\frac{\Opt^*_S}{r_S}+0.0139\frac{\Opt_L^*}{r_L})\\
&\ge\frac{1}{2}\cos(\frac{\theta}{2})\min\left\{\frac{0.0035}{r_S}+\frac{0.0139}{r_L}\right\}\Opt, \label{zonl}
\end{align}}
by parts ({\sf B1}) and ({\sf B3}) of Lemma~\ref{lem:l1}.
This completes the proof of the theorem.
\end{proof}
\medskip

\begin{customlem}{2}\label{lem:l1}
	The following inequalities hold:
\begin{itemize}
	\item[({\sf B1})] $\Opt^*\le\Opt_S^*+\Opt_L^*$;
	\item[({\sf B2})] $\Opt_F\ge\frac{\delta^2}{2}\Opt_L^*$;
	\item[({\sf B3})] $\Opt^*\ge\cos\frac{\theta}{2}\Opt$.
	\end{itemize}
\end{customlem}
\begin{proof}
({\sf B1}) Let $x^*$ be an optimal solution for \textsc{(CSP$_\textsc{R}$)}. Then $(x_k^*)_{k\in\cI_S}$ and $(x_k^*)_{k\in\cI_L}$ are feasible solutions for \textsc{(CSP$_\textsc{R}$[S])} and \textsc{(CSP$_\textsc{R}$[L])}, respectively.  Thus, $\sum_{k\in\cI_S}u_kx_k^*\le\Opt^*_S$ and $\sum_{k\in\cI_L}u_kx_k^*\le\Opt^*_L$, implying that $\Opt^*=\sum_{k\in\cI_S}u_kx_k^*+\sum_{k\in\cI_L}u_kx_k^*\le\Opt^*_S+\Opt^*_L$.

({\sf B2}) Let $x^*$ be an optimal solution for \textsc{(CSP$_\textsc{R}$[L])}. Then by Lemma~\ref{thm:delta}, $\sum_{k\in\cI_L:t \in T_k}x_k^*\le \frac{2}{\delta^2}$ for all $t\in\cT$. It follows that $\widetilde x=(\frac{\delta^2}{2}u_kx^*_k)_{k\in\cI_L}$ is a feasible solution for  \textsc{(FWMIS)}, and hence $\Opt_F\ge \sum_{k\in\cI_L}\widetilde x_k=\frac{\delta^2}{2}\sum_{k\in\cI_L}u_k x_k^*=\frac{\delta^2}{2}\Opt_L^*$.  

({\sf B3}) Let $x^*$ be an optimal solution for \textsc{(CSP$_\textsc{C}$)}. By Lemma \ref{lem:tb} below and using the fact that $x^*$ satisfies~\raf{mc1} and~\raf{mc3} we have

\begin{equation}
\displaystyle \cos \tfrac{\theta}{2}\cdot \sum_{k \in \cN:k\in T_k}x_k^* |S_{i,t}| \le \Big| \sum_{k \in \cN:k\in T_k}x_k^* S_{i,t}\Big | \le C_t \,\,\,\,\, \forall t \in \cT \, ,
\end{equation}

\noindent since  $\theta$ is restricted to be at most $\frac{\pi}{2}$. Thus, $x=(\cos \tfrac{\theta}{2}x_k^*)_{k\in\cN}$ is a feasible solution to \textsc{(CSP$_\textsc{R}$)}. This implies that $\Opt^*\ge\cos \tfrac{\theta}{2}\sum_{k\in\cN}u_kx_k^*=\cos \tfrac{\theta}{2}\Opt$.
\end{proof}

\medskip
\begin{customlem}{3}(\cite{KKCMZ16} )
	\label{lem:tb}
	Given a set of 2D vectors $\{d_i \in \RR^2\}_{i=1}^n$
	$$ \frac{\sum_{i=1}^n |d_i| }{\big| \sum_{i =1}^n d_i \big|} \le \sec\tfrac{\theta}{2},$$
	where $\theta$ is the maximum angle between any pair of vectors  $\{d_i \in \RR^2\}_{i=1}^n$ and $0 \le \theta \le \frac{\pi}{2}$.
\end{customlem}

\medskip
\begin{customlem}{4}
	\label{thm:delta}
	Fix a feasible solution $(x^*_k)_{k \in \mathcal{I}_L}$ to an instance of problem~\textsc{(CSP$_\textsc{R}$[L])}. Then under the NBA assumption, at any time instant $t\in\cT$, $\sum_{k\in\cI_L:t \in T_k} x_k^*\le \frac{2}{\delta^2}$.
\end{customlem}

\begin{proof}
	We use a similar argument to the one in Lemma 5.1 in \cite{Chakrabarti2002}, applied to the fractional problem. For each demand $k\in\cI_L$, identify a {\it bottleneck} time slot $t_k\in T_k$ such that $C_{t_k}=\min_{t\in T_k} C_t$.
	Fix a time slot $t\in\cT$. Let $\cI_L(t)$ be the set of all demands $k\in\cI_L$ such that $t\in T_k$. We partition the set of demands in $\cI_L(t)$ into two sets $\cI_L^\ell(t)$ and $\cI_L^r(t)$: a demand $k\in\cI_L(t)$ is in $\cI_L^\ell(t)$ 
	if it has its bottleneck time slot to the left of $t$ (including $t$), otherwise the demand is in $\cI_L^r(t)$. We show that 
	$\sum_{k\in\cI_L^\ell(t)}x_k^*\le \frac{1}{\delta^2}$ and a similar argument shows that $\sum_{k\in\cI_L^r(t)}x_k^*\le \frac{1}{\delta^2}$, which together imply the lemma.
	
	Let $A$ be the set of bottleneck slots for demands in $\cI_L^\ell(t)$ and let $t'$ be the {\it rightmost} slot in $A$.
	As $t'$ is a bottleneck slot for some $\delta$-large demand $j\in \cI_L^\ell(t)$ and by the NBA, we have $\delta C_{t'}\le|S_j|\le C_{\min}$, which implies that $C_{\min}\ge\delta C_{t'}$, and hence $|S_k|\ge\delta C_{t_k}\ge\delta C_{\min}\ge \delta^2 C_{t'}$ for any $\delta$-large demand $k\in\cI_L$. Since $t'$ is the rightmost time slot in $A$, $t'\in T_k$ for all demands $k\in\cI_L^\ell(t)$. Hence, form the above and the feasibility of $\widehat x$ to ~\textsc{(CSP$_\textsc{R}$[L])}, we get 
	$$
	\delta^2 C_{t'}\sum_{k\in\cI_L^\ell(t)} x_k^*\le\sum_{k\in\cI_L^\ell(t)}|S_k| x_k^*\le\sum_{k\in\cI_L(t')}|S_k|x^*_k\le C_{t'}.
	$$
	It follows that  $\sum_{k\in\cI_L^\ell(t)} x_k^*\le\frac{1}{\delta^2}.$
\end{proof}

\medskip
\begin{customlem}{5}
	\label{lem:l3}
	Fix a feasible solution $(\widetilde x_k)_{k \in \mathcal{I}_L}$ to an instance of problem~\textsc{(FWMIS)}. Then the integral solution $x=(x_k)_{k \in \mathcal{I}_L}$ obtained by setting $x_k=1$ with probability $\frac{\widetilde x_k}{2 u_k}$, followed by a correction step, has expected utility at least $\frac{1}{4}\sum_{k \in \mathcal{I}_L}\widetilde x_k$.
\end{customlem}
\begin{proof}
	For $k \in \mathcal{I}_L$, let $X_k\in\{0,1\}$ be a random variable indicating whether demand $k$ was selected initially, and $Y_k\in\{0,1\}$ be a random variable indicating whether demand $k$ survives the correction step. Then ${\mathbb P}[Y_k=1]=(1-{\mathbb P}[Y_k=0|X_k=1]){\mathbb P}[X_k=1]$. Let $t\in\cT$ be the time slot at which demand $k$ arrives. Note that ${\mathbb P}[Y_k=0|X_k=1]$ is the probability that there exists a demand $j\in\cI_L$  that arrived before $k$ such that $t\in T_j$ and $Y_k=1$. Thus, 
	\begin{align*}
	{\mathbb P}[Y_k=0|X_k=1]&\le\sum_{j\in\cI_L:t\in\ T_j}{\mathbb P}[X_j=1]\\
	&=\sum_{j\in\cI_L:t\in\ T_j}\frac{\widetilde{x}_j}{2 u_k}\le\frac{1}{2} \, ,
	\end{align*}
	    by the feasibility of $\widetilde x$ for~\textsc{(FWMIS)}. It follows that ${\mathbb P}[Y_k=1]\ge\frac{1}{2}\cdot\frac{\widetilde x_k}{2 u_k}$ and hence the expected utility of the integral solution obtained is $\sum_{k\in\cI_L}u_k{\mathbb P}[Y_k=1]\ge\frac{1}{4}\sum_{k \in \mathcal{I}_L}\widetilde x_k$.
\end{proof}
\medskip
\noindent\textbf{Remark:} It is worthwhile mentioning that to the best of our knowledge there are no known results in the literature concerning the lower bound of the competitive ratio of the online problem {\sc CSP$_\textsc{C}$}. From the arguments stated in~\cite{TCS-024}, however, it holds that $\mathcal{O}(\log(\frac{a^{\max}}{a^{\min}}))$ is indeed the best possible competitive ratio that could be achieved by any online algorithm for an instance of the online problem {\sc D} with only a single constraint, which will be matched by the achievable competitive ratio of our online algorithm in such case.

%% file: paper.bbl
\begin{thebibliography}{10}

\bibitem{TCS-024}
N.~Buchbinder and J.~S. Naor, ``The design of competitive online algorithms via
  a primal–dual approach,'' {\em Foundations and Trends in Theoretical
  Computer Science}, vol.~3, no.~2–3, pp.~93--263, 2009.

\bibitem{Borodin:1998}
A.~Borodin and R.~El-Yaniv, {\em Online computation and competitive analysis}.
\newblock Cambridge University Press Cambridge, 1998.

\bibitem{Lian13}
L.~Lu, J.~Tu, C.-K. Chau, M.~Chen, and X.~Lin, ``Online energy generation
  scheduling for microgrids with intermittent energy sources and
  co-generation,'' in {\em ACM SIGMETRICS}, 2013.

\bibitem{CZC16ESP}
C.-K. Chau, G.~Zhang, and M.~Chen, ``Cost minimizing online algorithms for
  energy storage management with worst-case guarantee,'' {\em to appear in IEEE
  Transactions on Smart Grid}, 2016.

\bibitem{CKA16DIA}
C.-K. Chau, M.~Khonji, and M.~Aftab, ``Online algorithms for information
  aggregation from distributed and correlated sources,'' {\em to appear in
  IEEE/ACM Transactions on Networking}, 2016.

\bibitem{YC13CKP}
L.~Yu and C.-K. Chau, ``Complex-demand knapsack problems and incentives in {AC}
  power systems,'' in {\em International Conference on Autonomous Agents and
  Multiagent Systems (AAMAS)}, 2013.

\bibitem{woeginger2000does}
G.~J. Woeginger, ``When does a dynamic programming formulation guarantee the
  existence of a fully polynomial time approximation scheme ({FPTAS})?,'' {\em
  INFORMS Journal on Computing}, vol.~12, no.~1, pp.~57--74, 2000.

\bibitem{CKM14}
C.-K. Chau, K.~Elbassioni, and M.~Khonji, ``Truthful mechanisms for
  combinatorial {AC} electric power allocation,'' in {\em International
  Conference on Autonomous Agents and Multiagent Systems (AAMAS)}, 2014.

\bibitem{CKM15}
C.-K. Chau, K.~Elbassioni, and M.~Khonji, ``Truthful mechanisms for
  combinatorial allocation of electric power in alternating current electric
  systems for smart grid,'' {\em to appear in ACM Transactions on Economics and
  Computation}, 2016.
\newblock http://arxiv.org/abs/1507.01762.

\bibitem{KKEC16CSPsss}
M.~Khonji, A.~Karapetyan, K.~Elbassioni, and C.-K. Chau, ``Complex-demand
  scheduling problem with application in smart grid,'' in {\em International
  Computing and Combinatorics Conference (COCOON)}, 2016.
\newblock http://arxiv.org/abs/1603.01786.

\bibitem{Marchetti-Spaccamela1995}
A.~Marchetti-Spaccamela and C.~Vercellis, ``Stochastic on-line knapsack
  problems,'' {\em Mathematical Programming}, vol.~68, no.~1, pp.~73--104,
  1995.

\bibitem{Chakrabarti2002}
A.~Chakrabarti, C.~Chekuri, A.~Gupta, and A.~Kumar, {\em Approximation
  Algorithms for the Unsplittable Flow Problem}, pp.~51--66.
\newblock Berlin, Heidelberg: Springer Berlin Heidelberg, 2002.

\bibitem{1339347}
P.~Korovesis, G.~Vokas, I.~Gonos, and F.~Topalis, ``Influence of large-scale
  installation of energy saving lamps on the line voltage distortion of a weak
  network supplied by photovoltaic station,'' {\em IEEE Transactions on Power
  Delivery}, vol.~19, pp.~1787--1793, Oct 2004.

\bibitem{baran1989placement}
M.~E. Baran and F.~F. Wu, ``Optimal capacitor placement on radial distribution
  systems,'' {\em IEEE Transactions on Power Delivery}, vol.~4, no.~1,
  pp.~725--734, 1989.

\bibitem{KCE2016OPF}
M.~Khonj, C.-K. Chau, and K.~Elbassioni, ``Optimal power flow with inelastic
  demands for demand response in radial distribution networks,'' {\em to appear
  in IEEE Transactions on Control of Network Systems}, 2016.
\newblock http://arxiv.org/abs/1601.02323.

\bibitem{KKCMZ16}
A.~Karapetyan, M.~Khonji, C.-K. Chau, K.~Elbassioni, and H.~Zeineldin,
  ``Efficient algorithm for scalable event-based demand response management in
  microgrids,'' {\em to appear in IEEE Transactions on Smart Grid}, 2016.
\newblock {http://arxiv.org/abs/1610.03002}.

\end{thebibliography}
